%
%
\catcode`@=11 
\font\titlefontB=cmssdc10 at 20pt
\catcode`@=12 
\def\skipaline{\vskip 12pt plus 1pt}
%
%
\magnification = \magstephalf
\hsize = 16.0 truecm
\vsize = 22.0 truecm 
\parindent=0.6 truecm 
\parskip = 2pt
\normalbaselineskip = 14pt plus 0.2pt minus 0.1pt
\baselineskip = \normalbaselineskip 
\nopagenumbers 
\topskip= 27pt plus2pt      
\voffset=2\baselineskip   
\hoffset= -0.01 truecm      
%
%
%
%
\def\downnormalfill{$\,\,\vrule depth4pt width0.4pt
\leaders\vrule depth 0pt height0.4pt\hfill\vrule depth4pt width0.4pt\,\,$}
\def\WT#1{\mathop{\vbox{\ialign{##\crcr\noalign{\kern3pt}
      \downnormalfill\crcr\noalign{\kern1.8pt\nointerlineskip}
      $\hfil\displaystyle{#1}\hfil$\crcr}}}\limits}

%
%
%
\def\tfract#1/#2{{\textstyle{\raise0.8pt\hbox{$\scriptstyle#1$}\over%
\hbox{\lower0.8pt\hbox{$\scriptstyle#2$}}}}}
\def\mezzo{\tfract 1/2 }

\def\quarto{\tfract 1/4}
\def\i2quarto{\tfract i^2/4}

%
%
%
%

\newread\epsffilein    
\newif\ifepsffileok    
\newif\ifepsfbbfound   
\newif\ifepsfverbose   
\newdimen\epsfxsize    
\newdimen\epsfysize    
\newdimen\epsftsize    
\newdimen\epsfrsize    
\newdimen\epsftmp      
\newdimen\pspoints     
\pspoints=1bp          
\epsfxsize=0pt         
\epsfysize=0pt         
\def\epsfbox#1{\global\def\epsfllx{72}\global\def\epsflly{72}%
   \global\def\epsfurx{540}\global\def\epsfury{720}%
   \def\lbracket{[}\def\testit{#1}\ifx\testit\lbracket
   \let\next=\epsfgetlitbb\else\let\next=\epsfnormal\fi\next{#1}}%
\def\epsfgetlitbb#1#2 #3 #4 #5]#6{\epsfgrab #2 #3 #4 #5 .\\%
   \epsfsetgraph{#6}}%
\def\epsfnormal#1{\epsfgetbb{#1}\epsfsetgraph{#1}}%
\def\epsfgetbb#1{%
%
%
\openin\epsffilein=#1
\ifeof\epsffilein\errmessage{I couldn't open #1, will ignore it}\else
%
%
   {\epsffileoktrue \chardef\other=12
    \def\do##1{\catcode`##1=\other}\dospecials \catcode`\ =10
    \loop
       \read\epsffilein to \epsffileline
       \ifeof\epsffilein\epsffileokfalse\else
%
%
          \expandafter\epsfaux\epsffileline:. \\%
       \fi
   \ifepsffileok\repeat
   \ifepsfbbfound\else
    \ifepsfverbose\message{No bounding box comment in #1; using defaults}\fi\fi
   }\closein\epsffilein\fi}%
%
%
\def\epsfsetgraph#1{%
   \epsfrsize=\epsfury\pspoints
   \advance\epsfrsize by-\epsflly\pspoints
   \epsftsize=\epsfurx\pspoints
   \advance\epsftsize by-\epsfllx\pspoints
%
%
   \epsfsize\epsftsize\epsfrsize
   \ifnum\epsfxsize=0 \ifnum\epsfysize=0
      \epsfxsize=\epsftsize \epsfysize=\epsfrsize
%
%
     \else\epsftmp=\epsftsize \divide\epsftmp\epsfrsize
       \epsfxsize=\epsfysize \multiply\epsfxsize\epsftmp
       \multiply\epsftmp\epsfrsize \advance\epsftsize-\epsftmp
       \epsftmp=\epsfysize
       \loop \advance\epsftsize\epsftsize \divide\epsftmp 2
       \ifnum\epsftmp>0
          \ifnum\epsftsize<\epsfrsize\else
             \advance\epsftsize-\epsfrsize \advance\epsfxsize\epsftmp \fi
       \repeat
     \fi
   \else\epsftmp=\epsfrsize \divide\epsftmp\epsftsize
     \epsfysize=\epsfxsize \multiply\epsfysize\epsftmp   
     \multiply\epsftmp\epsftsize \advance\epsfrsize-\epsftmp
     \epsftmp=\epsfxsize
     \loop \advance\epsfrsize\epsfrsize \divide\epsftmp 2
     \ifnum\epsftmp>0
        \ifnum\epsfrsize<\epsftsize\else
           \advance\epsfrsize-\epsftsize \advance\epsfysize\epsftmp \fi
     \repeat     
   \fi
%
%
   \ifepsfverbose\message{#1: width=\the\epsfxsize, height=\the\epsfysize}\fi
   \epsftmp=10\epsfxsize \divide\epsftmp\pspoints
   \vbox to\epsfysize{\vfil\hbox to\epsfxsize{%
      \special{illustration #1 scaled \number\epsfscale}
      \hfil}}%
\epsfxsize=0pt\epsfysize=0pt\epsfscale=1000 }%

%
%
{\catcode`\%=12 \global\let\epsfpercent=
%
%
\long\def\epsfaux#1#2:#3\\{\ifx#1\epsfpercent
   \def\testit{#2}\ifx\testit\epsfbblit
      \epsfgrab #3 . . . \\%
      \epsffileokfalse
      \global\epsfbbfoundtrue
   \fi\else\ifx#1\par\else\epsffileokfalse\fi\fi}%
%
%
\def\epsfgrab #1 #2 #3 #4 #5\\{%
   \global\def\epsfllx{#1}\ifx\epsfllx\empty
      \epsfgrab #2 #3 #4 #5 .\\\else
   \global\def\epsflly{#2}%
   \global\def\epsfurx{#3}\global\def\epsfury{#4}\fi}%
%
%
%
%

\newcount\epsfscale    
\newdimen\epsftmpp     
\newdimen\epsftmppp    
\newdimen\epsfM        
\newdimen\sppoints     
\epsfscale=1000        
\sppoints=1000sp       
\epsfM=1000\sppoints
%
\def\computescale#1#2{%
  \epsftmpp=#1 \epsftmppp=#2
  \epsftmp=\epsftmpp \divide\epsftmp\epsftmppp  
  \epsfscale=\epsfM \multiply\epsfscale\epsftmp 
  \multiply\epsftmp\epsftmppp                   
  \advance\epsftmpp-\epsftmp                    
  \epsftmp=\epsfM                               
  \loop \advance\epsftmpp\epsftmpp              
    \divide\epsftmp 2                           
    \ifnum\epsftmp>0
      \ifnum\epsftmpp<\epsftmppp\else           
        \advance\epsftmpp-\epsftmppp            
        \advance\epsfscale\epsftmp \fi          
  \repeat
  \divide\epsfscale\sppoints}
\def\epsfsize#1#2{%
  \ifnum\epsfscale=1000
    \ifnum\epsfxsize=0
      \ifnum\epsfysize=0
      \else \computescale{\epsfysize}{#2}
      \fi
    \else \computescale{\epsfxsize}{#1}
    \fi
  \else
    \epsfxsize=#1
    \divide\epsfxsize by 1000 \multiply\epsfxsize by \epsfscale
  \fi}

\epsfverbosetrue
%

\nopagenumbers

\rightline{IFUP-TH 2002-26}

\vskip 2.7 truecm 

\centerline {\titlefontB Gravitational deflection of light} 

\vskip 0.5 truecm


\vskip 0.5 truecm

\centerline {\titlefontB and helicity asymmetry} 

\vskip 2.7 truecm

\centerline  {\bf E. ~GUADAGNINI}

\vskip 1 truecm 

\centerline {Dipartimento di Fisica {\sl Enrico Fermi} dell'Universit\`a di Pisa}
\centerline {and INFN Sezione di Pisa} 
\centerline {Via F. Buonarroti, 2. ~ 56100 - PISA - Italy } 

\vskip 4 truecm 

\narrower {\noindent {\bf Abstract.}  ~The helicity modification of light polarization
which is induced by the gravitational deflection from a classical heavy rotating body, like
a star or a planet, is considered. The expression of the helicity asymmetry is derived;
this asymmetry signals the gravitationally induced spin transfer from the rotating body to the
scattered photons.}
 
\vskip 2 truecm

\noindent {\hrule height0.2pt width200pt depth0pt}

\medskip 

\noindent E-Mail:  ~guada@df.unipi.it \hfill

\vfill\eject
  
%
%
\count0 = 1   
%
%
%
\headline={\ifodd\pageno\rightheadline \else\leftheadline\fi} 
\def\rightheadline{{\hfil {\folio} \hfil }}
\def\leftheadline{{\hfil {\folio} \hfil}}
%
%

\noindent {\bf 1. The asymmetry.} ~Gyroscopic effects in gravity, like the Lens-Thirring  
 [1] and the Skrotskii effect [2] have been the subject of considerable research [3,4]. 
In this article  I shall consider a related issue; namely, the helicity modification of
the light polarization which is induced by the gravitational deflection from a classical heavy
rotating  source, like a star or a planet. The transition amplitude of the gravitational
scattering for the different polarization states of the photons will be computed. It turns out
that, because of the nonvanishing angular momentum of the classical source, the transition
probabilities which are associated with the two helicity states of the deflected photons may
differ. As a result, even if the incoming radiation is not polarized, the deflected photons
may have a nontrivial elliptic polarization corresponding to a  nonvanishing total intrinsic
spin.  In order to illustrate this phenomenon, let us consider for instance the light  
deflected from the Sun. As shown in Figure~1, suppose that the incoming flux of unpolarized
photons is directed as the angular momentum vector of the Sun.

\vskip 1.8 truecm

\centerline {\epsfscale=600 \epsfbox{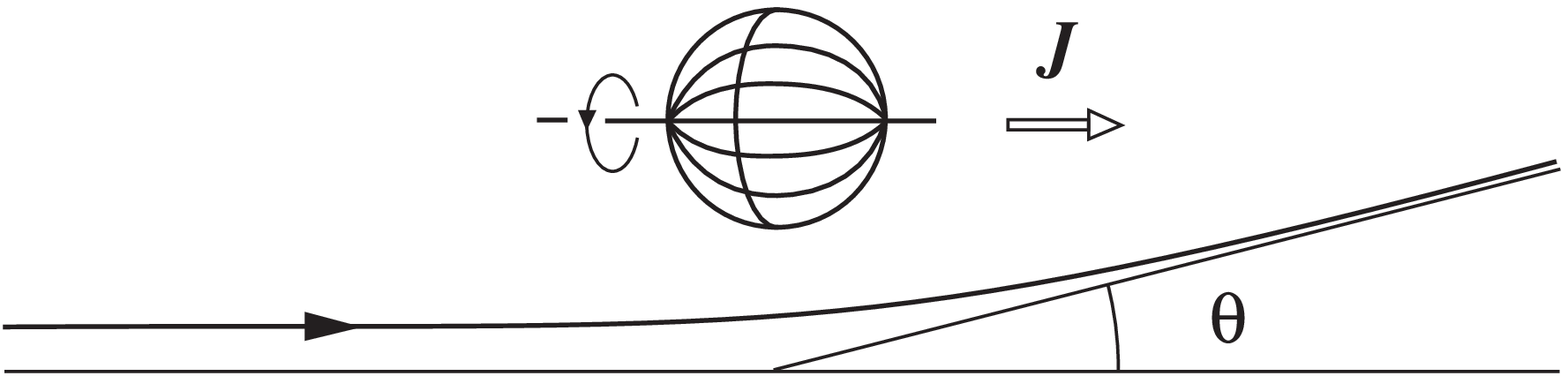}}

\vskip 0.7 truecm

\centerline {Figure~1. Light deflection.}

\vskip 1.1 truecm

\noindent Let $\, n_+ (\theta )\, $ ($\, n_- (\theta ) \, $)
be the number of  deflected photons with helicity $  \, +1 \, $ ($\, -1 \, $) at the scattering
angle $\, \theta \, $. Then, the prediction for the helicity asymmetry $ \, \chi ( \theta ) \,
$ is 
$$
\chi ( \theta ) \; \equiv \; {n_+ (\theta ) \, -\,  n_- (\theta ) \over n_+ (\theta ) \, +\, 
n_- (\theta )} \; \simeq \; 2 \, \pi \, {J \, \theta^2 \over M \, c \, \lambda } \quad , 
\eqno(1)
$$
where $ \, J \, $ denotes the magnitude of the angular momentum of the Sun, $ \, M \, $
represents its mass, and $ \, \lambda \, $ is the wave length of the electromagnetic
radiation.  By inserting the values $ \, J\simeq 10^{40}$~kg m$^2$/s , $\, \theta \simeq 4
\times 10^{-6} \, $, $ \, \lambda \simeq 5 \times 10^{-7} $~m , one finds 
$$
\chi \; \simeq \; 0.33 \; \% \quad , 
\eqno(2)
$$
which appears to be suitable for an experimental verification. The derivation of equation (1)
and a few comments on its possible applications are in order. 

\skipaline 

\noindent {\bf 2. The computation.} ~Let us consider  effective quantum
gravity [5,6], which provides the natural quantum field theory interpretation  of Einstein's
theory of gravitation. Effective quantum gravity is a phenomenological theory which can be
understood as the gravitational analogue  of the Fermi theory of the beta decay for weak
interactions. The spacetime metric $ \, g_{\mu \nu} (x) \, $ of general relativity is written 
as $ \, g_{\mu \nu} (x) = \eta_{\mu \nu } + h_{\mu \nu }(x) \, $, where $\, \eta_{\mu \nu} \, $
represents the Minkowski metric of flat spacetime; in cartesian coordinates $ \, \{ \, x^
\mu =(x^0 , \vec x \, ) \, \} $, $\, \eta_{\mu \nu} \, $ takes the standard diagonal form  $\,
{\rm diag.} ( \eta_{\mu \nu} ) =  ( +1 , -1 , -1 , -1 ) \, $. The effective quantum field
theory description of gravity is based  on a perturbative expansion in powers of $ \, h_{\mu
\nu } \, $. In the following computation, the standard conventions $  \, \hbar
= c = 1 \, $ will be adopted. 

\noindent $\bullet $  {\bf Gravitational couplings.} In the large distance limit and to first
order in $\, v/c \, $, the coupling of the fluctuation field $ \, h_{\mu \nu } (x) \, $ with a
classical heavy rotating body, which is subject to stationary conditions and is 
placed in position $ \, \vec r \, $, is given by the action term 
$$ \eqalign { 
S_{(1)} \; &= \; - \mezzo \int d^4 x \> \Theta^{\mu \nu} (\vec x\, ) \> h_{\mu \nu} (x) \cr 
&= \; - \mezzo \int d^4 x \> \left [ \, M \, h_{0 0} (x) \, + \, \epsilon^{ijk} \, J_i\,
\partial_j h_{0k} (x) \, \right ]  \, \delta^3 (\vec x - \vec r\, ) 
\quad , \cr }  
\eqno(3)
$$ 
where $\, M \, $ denotes the mass of the body and $ \, \{  \, J_i \, \} \, $ are the components
of its total angular momentum. The interaction of the metric fluctuation $
\, h_{\mu \nu } (x) \, $ with photons is described by the energy-momentum coupling  
$$ 
S_{(2)} \; = \; - \mezzo \int d^4 y \> T^{\mu \nu }( y ) \, h_{\mu \nu} (y) \qquad , 
\eqno(4)
$$
with 
$$
T^{\mu \nu }( y ) \; = \; F^{\mu \sigma }(y)\, F_\sigma^{ ~ \, \nu } (y) \, +\,  \quarto \,
\eta^{\mu \nu} \, F_{ \lambda \sigma} (y) \, F^{\lambda \sigma} (y) \quad . 
\eqno(5)
$$

\noindent $\bullet $  {\bf Gravitational scattering.} The gravitational
scattering of one photon from a classical heavy rotating body  is quite similar to 
the Rutherford scattering of one electron from a fixed Coulomb potential in electrodynamics. 
Suppose that, in the initial state, the photon has momentum $ \,
\vec p \, $ and polarization $ \, \alpha \, $; $ \, | \, {\rm in} \,
\rangle =  | \, \vec p \, , \alpha \rangle = a^+(\vec p , \alpha ) \, | 0 \rangle \, $ where $
\, a^+(\vec p , \alpha ) \, $ is the creation operator of the photon. Let $ \, \vec k \, $
and $ \, \beta \, $ be the final momentum and polarization of the photon. At the tree-level,
the amplitude $\,  A \, $ of this particular  ``newtonian scattering"  is given by 
$$
A \; = \;  \i2quarto \int d^4 x \, d^4y \> \Theta^{\lambda \sigma }(\vec x \, ) \,
{\WT{h_{\lambda \sigma} (x)\> h}}_{\mu \nu} (y) \, \langle \, \vec k \, , \beta \, | \, T^{\mu
\nu }( y  ) \,  | \, \vec p \, , \alpha \rangle \quad . 
\eqno (6)
$$
By using the Einstein-Hilbert action for the metric field $ \, g_{\mu \nu} (x) = \eta_{\mu \nu
} + h_{\mu \nu }(x) \, $, one can derive the expression of the graviton propagator which, in
Feynman gauge, takes the form 
$$
{\WT{h^{\mu \nu}(x)\> h}}_{\tau \sigma}(y) \; = \; i \, (16\pi G) \int {d^4p\over
(2\pi)^4}\, { e^{-ip(x-y)} \over p^2 \, + \, i \, \epsilon} \left(  
\delta^\mu_\tau \, \delta^\nu_\sigma \, + \, \delta^\mu_\sigma \, \delta^\nu_\tau 
\, - \, \eta^{\mu \nu} \, \eta_{\tau \sigma}   \right )  \quad . 
\eqno(7)
$$

\noindent $\bullet $  {\bf Transition amplitude.} Because of the invariance  under
time translations, energy is conserved and then $ \, |  \vec k \, | = |  \vec p \, | = p \,
$. The momentum transfer of the diffusion process is described by the vector $ \, ( \vec k -
\vec p \, ) \,  $ which has magnitude $ \, |   \vec k -  \vec p \, | = 2 p \sin \theta /2 \,
$ where $ \, \theta \, $ denotes the scattering angle. Let $ \, \vec \varepsilon_\alpha \, $
and $ \, \vec \varepsilon_\beta \, $ be the polarization vectors of the incoming and outcoming
photon respectively. The transition amplitude (6) can be written as 
$$
A \; = \; {i \, G \, M \, \delta ( |  \vec k \, | - |  \vec p \, | ) \over \pi \,  p \,
  |  \vec k -  \vec p \, |^2} \left ( B_1 \, + \, B_2 \right ) \quad , 
\eqno(8)
$$
where 
$$
B_1 \; = \; \mezzo |  \vec k +  \vec p \, |^2 \bigl ( \vec \varepsilon^{\, *}_\beta  \cdot
\vec \varepsilon_\alpha  \bigr ) \, - \, \bigl ( \vec p \cdot \vec \varepsilon^{\, *}_\beta
 \bigr ) \, \bigl ( \vec k \cdot \vec \varepsilon_\alpha  \bigr ) \quad , 
\eqno(9)
$$
$$
B_2 \; = \; {i \, p \over M}   \epsilon^{ijk} \, J_i \, \bigl ( \vec k - \vec p \, \bigr )_j
\biggl [ \bigl ( \vec k + \vec p  \, \bigr )_k \bigl ( \vec \varepsilon^{\, *}_\beta  \cdot
\vec \varepsilon_\alpha  \bigr ) \, - \, (\vec \varepsilon_\alpha )_k \, \bigl ( \vec p \cdot
\vec \varepsilon^{\, *}_\beta \bigr ) \, - \,  (\vec \varepsilon^{*}_\beta )_k \, \bigl ( \vec
k \cdot \vec \varepsilon_\alpha  \bigr )  \biggr ] \quad . 
\eqno(10)
$$
Let us introduce a basis of linear polarizations for the photons. The
scattering plane can be identified with the $ \, xy  $-plane; more precisely, the components
of the initial and final momenta are taken to be $ \, \vec p = ( p , 0 , 0 ) \, $ and 
$ \, \vec k = ( p \cos \theta , p \sin \theta , 0 ) \, $. Thus, one can put 
$$
\vec \varepsilon_\alpha (1) \; = \; ( 0 , 1 , 0 ) \; , \; 
\vec \varepsilon_\alpha (2) \; = \; ( 0 , 0 , 1 ) \; , \;
\vec \varepsilon_\beta (1) \; = \; ( - \sin \theta , \cos \theta , 0 ) \; , \; 
\vec \varepsilon_\beta (2) \; = \; ( 0 , 0 , 1 ) \; .
$$
The amplitude can then be written in matrix form with respect to the polarization states 
$$
\left [ \matrix { (1)_\alpha \otimes (1)_\beta & (1)_\alpha \otimes (2)_\beta \cr
(2)_\alpha \otimes (1)_\beta & (2)_\alpha \otimes (2)_\beta \cr } \right ] \quad . 
$$
One finds
$$
A \; = \; {i \, G \, M \,  \delta ( |  \vec k \, | - |  \vec p \, | ) \,  \cos \theta /2 \over
2 \, \pi \, p \, \sin^2 \theta /2 } \left [ \matrix { 
\cos \theta /2 - i {2 p J_3  \over M} \sin \theta /2 & 
-i {2 p {\widetilde J}  \over M} \sin^2 \theta /2 \cr 
i {2 p {\widetilde J} \over M} \sin^2 \theta /2 & 
 \cos \theta /2 - i {2 p J_3  \over M} \sin \theta /2 \cr } \right ] 
\eqno(11)
$$
where  $ \, J_3 \, $ is the component of the angular momentum of the source which is
orthogonal to the scattering plane whereas $\, \widetilde J \, $, given by
$$
\widetilde J \; = \; J_1 \, \cos \theta /2 \, + \,  J_2 \, \sin \theta /2  \quad ,
\eqno(12)
$$ 
is the component of the angular momentum  which belongs to the scattering plane and is
orthogonal to $ \, ( \vec k - \vec p \, ) \, $.  Note that the $ \, J_3 $-contribution  to
the amplitude is proportional to the identity matrix; so, the component $ \, J_3 \, $ does not
modify the polarization state of the photons. 

Now, the newtonian scattering we wish to analyze is similar but not equal to the Coulomb
scattering of electrons. In the gravitational deflection of light from a star or a planet, the
typical wavelength of the photon is very small if compared to the distance of the photon
from the source or the size of the source.  In practice, the wave packet associated with
the orbital state of the photon moves in an ``almost constant" gravitational potential.  
In the limiting case of an ``exactly constant" external potential, the momentum transfer 
would vanish. Consequently, in order to extract from expression (11) the  part $ \, A_{\rm rel}
\, $  of the transition amplitude which is relevant for light deflection, one must consider
the  $ \, ( \vec k - \vec p \, ) \rightarrow 0\,  $ limit.  The vector $ \, ( \vec k - \vec p
\, ) \,  $ has magnitude $ \, | \vec k -  \vec p \, | = 2 p \sin \theta /2 \, $; since  energy
is fixed, $ \, p \, $  is fixed and thus the  $ \, ( \vec k - \vec p \, ) \rightarrow 0\,  $
limit is equivalent to the  $ \, \theta \rightarrow 0 \, $ limit.  The leading term of each
matrix element of the amplitude (11) in the formal $ \, \theta \rightarrow 0 \, $ limit gives
$$
A_{\rm rel} \; = \; {i \, 4 \, G \, M \,  \delta ( |  \vec k \, | - |  \vec p \, | )  \over
2 \, \pi \, p \,  \theta^2} \left [ \matrix { 
1 &  -i  \, J_1 \, p \, \theta^2 / 2  M    \cr 
i  \, J_1 \, p \, \theta^2 / 2  M  &  1 \cr } \right ] \quad .
\eqno(13)
$$
It will now be assumed that expression (13) represents the relevant transition amplitude for
the gravitational deflection of light. The component $ \, J_1 \, $ of angular momentum which
enters equation (13) simply denotes the component of angular momentum which is parallel to
the incoming momentum $ \, \vec p \, $. Thus, one really has $ \, J_1 =  ( \vec
J \cdot \vec p \, ) \, / \, | \vec p \, |  \, $. 

As a first check, let us consider the differential cross section; in the $ \, \theta
\rightarrow 0 \, $ limit, equation (13) implies
$$
{d \sigma \over d \Omega }  \; = \; {16 \, G^2 \, M^2 \over \theta^4 } \quad . 
\eqno(14)
$$
On the other hand, in the semiclassical limit the cross section takes the form
$$
{d \sigma \over d \Omega }  \; \simeq \;  { b  \over  \theta } \, \left | { d b \over d \theta
} \right |\quad , 
\eqno(15)
$$
where $ \, b \, $ is the impact parameter. By comparing expressions (14) and (15), one
finds 
$$
\theta \; \simeq \; {4 \, G \, M \over b } \quad ,  
\eqno(16)
$$
which is in agreement with the prediction based on classical arguments [7]. 

\noindent $\bullet $  {\bf Helicity states.} Let us now concentrate on photon polarizations.
According to the result (13), if the polarization state of the incoming photon is described by
$\,  \vec \varepsilon_\alpha (1)  =  ( 0 , 1 , 0 ) \, $ then, in the $ \, \theta \rightarrow 0
\, $ limit, the polarization vector
$ \, \vec \varepsilon \, $ of the deflected photon is 
$$
\vec \varepsilon \; \simeq \; \vec \varepsilon_\beta (1) \, + \, i  \left (  \vec
J \cdot \vec p  \, \, \theta^2 / 2  M \right ) \, \vec \varepsilon_\beta (2) \quad ,
\eqno(17)
$$
which corresponds to elliptic polarization. This phenomenon differs from the Skrotskii effect
which only concerns the rotation angle of the transported electric field direction.  For the
outcoming photons, let us introduce the polarization vectors $\, \vec \varepsilon_\beta (\pm )
\, $ corresponding to polarization states with $\, \pm 1 \, $ helicity
$$
\vec \varepsilon_\beta (\pm ) \; = \; {1\over \sqrt 2 } \left ( \, \vec \varepsilon_\beta (1)
\, \pm  \, i \,  \vec \varepsilon_\beta (2) \,  \right ) \quad . 
\eqno(18)
$$
Then relation (17) takes the form 
$$
\vec \varepsilon \; \simeq \; {1\over  \sqrt 2 } \, \left [ \, 1 \, + \,    \left (  \vec J
\cdot \vec p  \, \, \theta^2 / 2  M \right )  \, \right ] \, \vec \varepsilon_\beta ( + )
\, + \,  {1\over  \sqrt 2 } \, \left [ 1 \, - \,   \left (  \vec J \cdot \vec p  \, \,
\theta^2 / 2  M \right ) \, \right ]  \, \vec \varepsilon_\beta ( - ) \quad .
\eqno(19)
$$
Assuming $ \,\left (  \vec J \cdot \vec p  \, \, \theta^2 / 2  M \right )   \ll 1 \, $, the
probability  $ \, w (\pm) \, $ of detecting a scattered photon with helicity $ \, \pm 1
\, $ is 
$$
w(\pm ) \; \simeq \; \mezzo \, \pm \, \left (  \vec J \cdot \vec p  \, \, \theta^2 / 2 
M \right )  \quad . 
\eqno(20)
$$
Consequently, when the initial polarization is described by $\,  \vec \varepsilon_\alpha (1)
\, $,  one has
$$
w(+) \, - \, w(-) \; \simeq \; \left (  \vec J \cdot \vec p  \, \, \theta^2 /   M
\right )  \quad . 
\eqno(21)
$$
It is easy to verify that, when the initial polarization is described by $\,  \vec
\varepsilon_\alpha (2) =(0,0,1) \, $,  equation (21) is still valid. This concludes the
derivation of the helicity asymmetry expression (1) for the gravitational deflection of light. 

\skipaline 

\noindent {\bf 3. Gravitational spin transfer.} ~The helicity asymmetry is originated by a
nontrivial angular momentum of the deflecting body. A simple argument explains the
structure of expression (1). Consider the coupling (3) of the classical source with the 
metric fluctuation $ \, h_{\mu \nu} \, $.  With respect to  the gravitational mass term $ \, M
h_{00}  \, $, which does not induce helicity modifications, the angular momentum term $ \,
\epsilon^{ijk}  J_i \partial_j h_{0k} \, $ has strength $ \, J / M \, $ times the magnitude 
$ \, p \, \theta \, $ of the momentum transfer which corresponds to the spatial derivative
acting on $ \,  h_{0k} \, $. In standard $ \, \hbar = c = 1 \, $ units, the value $ \, p \, $
of the momentum of the photon equals $ \, 2 \pi / \lambda \, $. Consequently,
the non-diagonal matrix elements of the transition amplitude must be  of order $ \, 2
\pi J \theta / M \lambda \, $ (at least) with respect to the diagonal elements. The actual
computation shows that non-diagonal matrix elements contain an extra multiplicative $ \,
\theta \, $ factor.  Because of the spatial derivative acting on $ \,  h_{0k} \, $, the
contribution of the lagrangian term $ \, \epsilon^{ijk}  J_i \partial_j h_{0k} \, $ to the
transition amplitude  must contain the imaginary unit $ \, i = \sqrt {-1 \, } \, $ with respect
to the remaining part, and this implies the occurrence of helicity flips during the
scattering.   Finally, the sign of the asymmetry is fixed by consistency; in agreement with
the results of the computations, the spin excess of the deflected photons must be directed as
the angular momentum vector of the  source.  

The phenomenon associated with equation (1) can be understood as a  gravitationally induced
spin transfer from a classical rotating body to the scattered photons.  Even if expression
(1) does not contain $ \, \hbar \, $, it is not obvious how to explain this spin transfer in
purely classical terms.  Formula (1) has been obtained by using the simplest modelling of the
gravitational scattering and can be improved by taking into account the particular details 
of real macroscopic bodies.  Finally, it is important to note that the asymmetry (1)
corresponds to the vacuum optical activity; in practice, ordinary electromagnetic effects must
also be considered. 

In principle, the gravitationally induced spin transfer from a classical rotating body to
photons appears to be one of the simplest tests of the existence of gyroscopic effects in
gravity. This does not necessarily mean that the actual experimental  verification of this
phenomenon will be an easy task. For example, electromagnetic contributions to the helicity
asymmetry should be clearly identified. In the context of gravitational lensing, the
gravitational helicity asymmetry could be used to determine the magnitude of the angular
momentum of rotating galaxies. 

Aspects of the light propagation in various spacetimes have been discussed for instance by
L.~Blanchet, S.~Kopeikin and G.~Schafer, M.P.~Haung and C.~Lammerzahl, and Wei-Tou Ni in 
ref.[4]; additional material can be found in [8,-,11] and referenced quoted therein. Recently,
classical gravitomagnetic effects have been considered also in [12,13,14]. 

\skipaline

I wish to thank Steven Shore for useful discussions.

\vskip 2 truecm 

{\centerline {\bf References}}

\medskip 

\item {[1]} J.~Lense and H.~Tirring, Physik Zeitschr. 19 (1918) 156.

\item {[2]}  G.V.~Skrotskii, Doklady Akademii Nauk SSSR 114 (1957) 73.

\item {[3]} C.M.~Will, {\sl Theory and experiment in gravitational physics,}  Cambridge
University Press (New York, 1993).

\item {[4]} C.~Lammerzahl, C.W.F.~Everitt and F.W.~Hehl, {\sl Gyros, Clocks,
Interferometers...: Testing Relativistic Gravity in Space,} Lecture Notes in Physics 562,
Springer Verlag (Berlin, 2001). 

\item {[5]} R.P.~Feynman, F.B.~Morinigo and W.G.~Wagner, {\sl Feynman lectures on
gravitation,}  Penguin Books (Clays Ldt, St Ives plc, England, 1999).

\item {[6]} G.~'t~Hooft and M.~Veltman, Ann. Inst. Henry Poincar\'e 20 (1974) 69.

\item {[7]} C.W.~Misner, K.S.~Thorne and J.A.~Wheeler, {\sl Gravitation,} W.H.~Freeman and
Company (New York, 1997).

\item {[8]} A.~Stebbins, Astrophys. J. 327 (1988) 584.

\item {[9]} S.~Seitz, P.~Schneider and J.~Ehlers, preprint astro-ph/9403056.

\item {[10]} E.~Audit and J.F.L.~Simmons, Class. Quant. Grav. 11 (1994) 2345. 

\item {[11]} S.M.~Kopeikin and G.~Schafer, Phys. Rev. D60 (1999) 124002.

\item {[12]} S.~Kopeikin and B.~Mashhoon, Phys. Rev. D65 (2002) 064025. 

\item {[13]} A.~Tartaglia, preprint gr-qc/0201014.

\item {[14]} L.~Iorio, preprint gr-qc/0207005.

\vfill\eject
\bye 
\end